\documentclass[a4paper,twocolumn]{article}

\usepackage[english]{babel}
\usepackage[utf8x]{inputenc}
\usepackage{authblk}
\usepackage{amsmath}
\usepackage{booktabs}
\usepackage{graphicx}
\usepackage{xurl} 
\usepackage{breakurl}
\usepackage{sectsty}
\usepackage{parskip}
\usepackage[table,xcdraw]{xcolor}
\usepackage[colorinlistoftodos]{todonotes}
\usepackage[font={small}]{caption}
\pdfoutput=1

\title{Decentralized Basic Income: Creating Wealth with On-Chain Staking and Fixed-Rate Protocols}

\author[1,*]{Hakwan Lau}
\author[2,$\dagger$]{Stephen Tse}
\affil[1]{Center for Brain Science, Riken Institute, Japan}
\affil[2]{Harmony.ONE}
\affil[*]{Correspondence to: hakwan@gmail.com}
\affil[$\dagger$]{Correspondence to: s@harmony.one}

\date{\vspace{-0.2ex}} %5 previously set by Taylor 
\begin{document}
\maketitle

\makeatletter
\renewcommand*{\@seccntformat}[1]{\csname the#1\endcsname\hspace{0.2cm}}
\makeatother

\sectionfont{\fontsize{12}{14}\selectfont}
\begin{abstract}
%\smallskip
\vspace*{-0.6em}

\textbf{In this review, we evaluate the mechanisms behind the decentralized finance protocols for generating stable, passive income. Currently, some savings interest rates can be as high as 20\% annually, payable in traditional currency values such as US dollars. Therefore, one can benefit from the growth of the cryptocurrency markets, with minimal exposure to their volatility risks. We aim to explain the rationale behind these savings products in simple terms. The key here is that asset deposits in cryptocurrency ecosystems are of intrinsic economic value, as they facilitate network consensus mechanisms and automated marketplaces. Therefore, savings in cryptocurrency are associated with some unique advantages unavailable in traditional financial systems. We will go through the implementations of how savings can be channeled into the staking deposits in Proof-of-Stake (PoS) protocols, through fixed-rate lending protocols, and staking derivative tokens. We will discuss potential pitfalls, assess how these protocols may behave in market cycles, as well as suggest areas for further research and development. We end by discussing the notion of decentralized basic income – analogous to universal basic income but guaranteed by financial products on blockchains instead of public policies.
}
\end{abstract}

\bigskip

\section{Background}

As of year 2021, the interest rates for savings in traditional ‘fiat’ currencies such as the US dollar and the Japanese yen are at historic lows, at around 0.07\% and 0.01\% respectively\cite{gordon_2021}. Because of the large national debts held by these major economies, some may anticipate that interest rates may remain relatively low for the near future. At low interest rates, the traditionally lauded act of regular household savings becomes less rewarding. This may have a particularly strong impact on aging populations which rely on saved money as pension income, for the long-term sustenance of quality of life after retirement\cite{bindseil_2018}.

Meanwhile, during the pandemic triggered by COVID-19, major governments have printed large amounts of money to support their economic stimulation efforts \cite{wolverson_2018}. This has led some to worry that inflation, i.e. the rising of nominal prices of goods, will eventually become inevitable. Although modern economists have disputed the basis of this worry \cite{FRB_2003,kelton_2020,palley2015money}, historically this has been the consequences of excessive increases in circulating currency, such as in the cases of Germany in 1923 \cite{bresciani2013economics} and Argentina in 1989 \cite{kiguel1995inflation}. To many, these dramatic cases likely reinforces the pessimistic narrative about inflation as a source of potential concern. As of 2021 May, the US inflation rate for consumer goods was reported to be 5\% for the past annual period, considerably higher than anticipated \cite{guilford_2021}.

These may be some of the reasons why, somewhat paradoxically, despite the obviously negative economic impact of COVID-19, stocks and real estate prices have generally gone up since March 2020 \cite{real_estate,domm_2020}. As investors lose confidence in the future value of savings in fiat currencies, they look for options that would give higher expected returns. Accordingly, the cryptocurrency market has also flourished during this time \cite{jabotinsky_2021,demir2020relationship,sarkodie2021covid}. 

However, this puts the common household savers into an unfavorable position. To the socio-economic groups who may have the most need to hedge against possible inflation, fixed-term saving and government bonds have historically been the major go-to options for the steady accumulation of wealth. To the average saver, the stock market may seem like a rather complex and potentially risky place. While real estate may be perceived as a relatively stable asset, the threshold for investment entry is higher. Above all, cryptocurrency remains very far from a mainstream investment option. Importantly, the perceived risk is in fact supported by the evidently high volatility of the market.

But what if one can benefit from the staggering growth of the cryptocurrency market without ever ‘investing’ into it? That is, what if one can sidestep the financial risks directly associated with the acknowledged volatility of the market, and yet generate stable income via interest rates on the order of several hundred times higher (i.e. 20\% annual rate) than what the current traditional financial market can offer? What if this can happen even when the market is contracting instead of growing, at least to some extent?

This possibility is obviously of strategic concern for those who are already invested into cryptocurrency technology, who see mass adoption as a desirable and necessary next step for the development of the industry. But more importantly, there is also a moral argument behind this consideration, even for the technologically uninitiated. Given the current economic context, the widening gap between the rich and the poor may increasingly turn into a gap between savvy investors and traditional savers. The former group can afford taking higher risks due to larger initial capital, while the latter may lack the means and the know-how to truly benefit from anything other than the anemic fiat saving rates. 

Fortunately, certain existing cryptocurrency protocols may be able to help us bridge this gap. To anticipate, the general rationale behind is simple: in many modern blockchain systems the integrity of the ledger is supported by the Proof-of-Stake (PoS) consensus mechanism \cite{king2012ppcoin,nair2021evaluation}, rather than the Proof of Work (PoW) protocol used in e.g. Bitcoin. This means that to validate a transaction, one does not need to dedicate an excessive amount of energy and computational power to solve a cryptographic puzzle, to ‘mine’ a coin. Instead, in PoS networks, a validator needs to show that one holds a sufficient amount of the currency (i.e. stake) within the system such that it would be against one’s interest to act maliciously. By lending money to validators, one is allowing them to qualify to do their jobs. This process, sometimes called ‘staking’, means that putting money down for a fixed term can be of intrinsic economic value, as staking is key to the financial efficiency, security, and guaranteed integrity of the blockchain bookkeeping system. As long as a PoS ecosystem is able to derive utility and value with decentralized finance applications, e.g. for trading, there will be demand for staking. As such, savers can be rewarded in a sustainable way, as they support the generation of financial services and products provided by the PoS network - not entirely unlike why bankers and corporate lawyers are paid good salaries in the traditional financial system, even during downturns. 

\section{Stablecoins and Lending Markets}

Besides the existence of PoS blockchain systems, the protocols for the generation of stable interest assumes the availability of several instruments and decentralized finance (DeFi) products \cite{werner2021sok}. The most basic of these include stable currencies (sometimes called stable coins) pegged to real-world fiat currencies. Examples include Tether (USDT) \cite{tether}, MakerDAO’s DAI \cite{makerdao}, Binance’s BUSD \cite{binance}, Terra’s UST \cite{kwon}. These are all pegged to the US dollar. There are also stablecoins for other currencies, such as Korean won \cite{kereiakes2019terra}, British pound \cite{london_block}, etc. The different mechanisms for how the peg is achieved have been reviewed elsewhere \cite{moin2020sok,clark2019sok}. One interesting fact is that they do not necessarily have to be backed by real-world fiat assets. Some are backed by other cryptocurrencies, yet others maintain their value by algorithmically adjusting the supply. Accordingly, they come with varying degrees of absolute stability. Nevertheless, overall, for the major stablecoins including the ones mentioned above, the stabilization mechanisms have been shown to be generally effective over the past couple of years, withstanding considerable market volatility. Also, insurance options are now available to mitigate the risks of unpegging \cite{mutual,unleashed}. 

Stablecoins are important for our purposes; for the stable interest rates to be meaningful to the typical household savers, both the savings and the interest have to be in terms of everyday currencies. With them, one can already earn savings interest without exposing oneself directly to the volatility of the cryptocurrency market. Protocols like COMPOUND \cite{leshner2019compound}, Maker \cite{makerdao}, and AAVE \cite{aave} allow loans to be provided to borrowers who put down cryptocurrencies as collateral. Savers can deposit money into these protocols to form a pool for the loans. Because borrowers have to pay interest on their loans, part of that translates into the interest to be paid back to the savers in return. These interest rates are determined by market mechanisms \cite{bartoletti2020sok}, reflecting the expected values of future assets.

In part due to the semi-anonymized nature of financial engagement with cryptocurrency, with few exceptions \cite{trusttoken}, long-term loans tend to be overcollateralized, meaning that one has to put down a higher amount of ‘collateral’ in order to borrow. This overcollateralization ratio is typically up to 150\% to 200\%, meaning that for every \$1 borrowed one has to put down as collateralization something worth up to \$1.5 or \$2. This way, if a borrower defaults, the lender would be protected. Also, the collateral itself is typically in a cryptocurrency that fluctuates in price. Should the worth of the collateral diminishes so that the expected overcollateralization ratio cannot be maintained, the lender can seize the collateral (in accordance with prior agreement) and sell it immediately to avoid a loss. With overcollateralization, even slippage during a flash crash should not lead to a loss on the lender’s part. If the process of liquidation is achieved more efficiently and quickly, some lenders can even afford to offer a lower overcollateralization rate e.g. at just 110\% \cite{liquidity}.  Accordingly, ‘bad debt’ as we traditionally understand it does not really occur as such in decentralized finance. Given the nature of contracts in advanced blockchain systems (known as Smart Contracts \cite{bartoletti2017empirical, buterin2014next}), theoretically they are completely binding and frictionless to execute. 

In terms of economic incentives, some borrowers may be willing to put down collateral because they expect the value of them to go up in the future. So they do not want to sell the collateral at the current price. And yet, while ‘holding’ the cryptocurrency for the longer term, if the asset is deposited as collateral, the loan gives them instant liquidity, which may allow them to reinvest further into the market via leverage. This is part of the reason why they are willing to pay for the loan interest. In this sense, at least part of the savings interest ultimately comes from the expected growth of the cryptocurrency market, even though the savers do not have to partake into the market speculation directly in order to enjoy the interest as income. 

However, as of the time of writing, the interest rate for borrowing and lending tends not to exceed 10\% annual rate \cite{cryptolending}. This is already far higher than current fiat interest rates. In fact, this is on par with, or just outperforms in some cases, most other major forms of low-risk investments, like national bonds, pension accounts, etc.  And if one is to be paid back in stablecoins, the risks involved are likewise relatively small. However, one caveat is that it is unclear if borrowers would be willing to pay high interest during a bear market. Importantly, as we will discuss below, it may also be possible to generate stable interests at as high as 20\% annual rate, by capitalizing on other sources of revenue besides anticipated market growth.

\section{Fixed-Rate Lending Protocols and Derivatives}

One other issue in treating lending interest as stable income is that until recently, in decentralized finance, borrowing and lending interest rates tended to fluctuate flexibly according to market conditions. However, fixed-rate lending protocols are now available \cite{messari_fixed}. The key idea here is that we can trade the value of a loan on the market. That is, if a token is to give its holder the right to be paid \$100, as a settlement of a loan, at a certain future date, then the token itself could be traded at a certain price. Assuming one generally prefers money right now over money in the future, one could probably only be willing to buy this token at a discount. How deep that discount is, in turn, should depend on when the loan matures. For loans that are to be settled further in the future, buyers may be willing to buy such a token only at a deeper discount. That is to say, the interest for longer-term loans should be higher. By setting up automated markets for these tokens \cite{robinson2020yield}, one can work out the general interest rates for loans for different durations - what is sometimes called a yield curve \cite{messari_yield}. This curve is not entirely fixed in the long run. But it can allow relatively stable predictions to be made within a certain timeframe. 

The token described above is essentially a bond. Specifically, it is what is called a zero-coupon bond, meaning that the bond itself generates no interests or other additional benefits; the attractiveness is just that it may be bought at a current price lower than the eventual loan settlement. 

In traditional financial terms, bonds are a kind of derivatives, which are simply contracts that ‘derive’ their value based on some other asset. Some other types of derivatives include options for buying a certain asset at a certain price X at a certain future date. The holder may exercise the option only if the price turns out to be favorable. This could be useful, for example, for hedging purposes. That is, given that a party is obligated to fulfil a transaction in the future in a particular currency, such as to pay up for some consumed services. The party could end up suffering an unexpected loss if the price of the currency goes up. By holding the option token mentioned above, the party can be sure that the actual total cost would not exceed a certain point; if the price of the currency goes up above expectation, one can then exercise the option to buy the currency at the agreed price X, making a profit that would exactly compensate for the loss. 

Notably, the derivatives market in the traditional financial world is large, in which the most traded type of derivatives are interest rate swaps. The trade volume for these alone are on the order of 10 times larger than the stock market \cite{bond_market}. These instruments allow two parties to trade on the future yield of an underlying asset. This can be useful for turning floating interest rate incomes into a fixed-rate income, or vice versa. For example, if I have lent out money to someone who is to pay me back at whatever current fiat interest rate is, I may be able to find a third party to swap my interest income into a fixed rate. The third party may demand a lump sum, or may only offer a relatively low fixed rate. But if the deal is made, the third party will absorb the variability in future changes in fiat interest rates. 

We will return to interest rate swaps in section 8. For now, what is important is that this sets the context for a type of derivative that is unique to decentralized finance, and is key to the problem of generating stable income. 

\section{Staking Derivatives}

In a PoS network, staking derivatives are tokens that can be generated when one lends money to validators to allow them to qualify for the job of approving transactions \cite{konstantopoulos,chitra2020stake}. Without such tokens, there will be a number of risks involved in staking. The first is due to the fact that staking generally requires locking in the staked currency for a fixed period, typically up to a few weeks, during which the validators are considered qualified for the role (by demonstrating that they have enough stake). If the currency drops in value over this time, the lender may not be able to ‘unstake’ fast enough, i.e. to extract the money back from the validator, to stop the losses. The second is that if the validator fails to do the job correctly, such as not being present frequently enough to approve new transactions on time, there will be penalties, i.e. part of the staked currency will be ‘slashed’. 

By having staking derivatives, it addresses the first problem by making the staked value essentially liquid. The staking derivative can be the contract allowing whomever holding the token to claim the staked currency, as the time matures. Therefore, by trading the staking derivative, one is essentially selling off the loan to the validator to someone else, in exchange for instant liquidity. As different people may have different speculations on the future value of the staked currency, general market mechanisms will determine its current price. Depending on the nature of the contract, the risk of slashing can also be taken into account.

Currently, not all PoS networks offer staking derivatives, at least not natively. However, the concept behind is straightforward and attractive, and we expect implementation and adoption to occur widely in the near future. As we will see in the next section, most relevant to the protocols for the generation of high stable interest is the fact that these staking derivatives can be used as collaterals for borrowing money. 

\section{Staking Fees as Stable Interest}

Because staking is essential for a PoS ledger system to maintain its integrity, staking fees on the order of 10\% annual rate or higher are typically offered to incentivize stakers \cite{major_2021}. Importantly, these fees are relatively stable. Depending on how the derivative contract is set up, the fee can go entirely towards the holder of the relevant staking derivative. In the Anchor protocol \cite{anchor_protocol}, one can put down staking derivatives as collateral, for borrowing money in a stable currency. Instead of forfeiting all the staking fees, the borrower can continue to earn a portion of it. The exact percentage of this split is not a constant, as we will explain in the next section. But for illustration we can assume that this split will be 1:5, meaning the borrower gets to retain 1/6 of the staking fees earned. That means they give up 5/6 of the staking fees yield, in exchange of the instant liquidity provided by having the loan. (Nominally, the ‘splitting’ is more complex in the Anchor protocol, in that the borrower pays an interest, and receives certain tokens of value, ANC, in return. But the sum total of the transaction is equivalent to losing part but not all of the staking fees earned, and for simplicity we will continue to conceptualize and present it this way.)

Like other lending protocols discussed in section 2, the Anchor protocol thereby also functions a bit like a bank in a traditional financial system. To accumulate a pool of deposited money for lending to borrowers, it accepts savings in a stable currency from savers. Because it takes a split of the staking fees from the borrower’s collaterals, that can be used to contribute towards the interests to be paid to the savers. Using the example above of a 1:5 split of the staking fees between the borrowers and the protocol, at an overcollateralization ratio of 200\%, the maximum affordable interest rate can be as high as 167\% the staking fee itself; each dollar borrowed attract twice as much worth of collateral, which in turn attracts 5/6 of the staking fees per collateral unit (5/6 x 2 = 167\%). This is why the Anchor protocol can offer such competitive interest rates, currently at about 20\% annual rate, given that the staking yield is at 12\% on the native LUNA network \cite{kereiakes2019terra} for staking (12\% x 167\% ≈ 20\%). In other words, it exploits a source of income that is not available in traditional financial systems, and is unique to PoS networks: staking fees.

But if the interest ultimately comes from staking fees, why do savers not stake the money themselves and earn the staking fees directly? Or, assuming some savers may lack the technical knowhow and familiarity with cryptocurrencies to engage in such activity, why does the savings protocol (such as Anchor) not directly stake the savers’ money on their behalf, and use the generated staking fees to pay the interest? Why do we need to involve the borrowers at all? 

There are several advantages in doing so. The first is that staking is inherently risky, as explained briefly in the last section. During the period of staking, the value of the staked currency may fluctuate. Also, if the savings protocol collects money from many savers, and proceeds to stake the entire pool of money, this may also create an unhealthy situation for the staked network. Ultimately, the PoS consensus mechanisms work well when there are many different independent stakeholders, all incentivized to make honest and reliable decisions, none of whom are ‘too big to fail’. If all the validators are chosen and supported by a single source of money, the security of the network could be compromised  \cite{chitra2020stake}. By having borrowers who independently make their staking decisions, and contribute staking derivatives voluntarily as collaterals, the risks and control are both diluted over many individuals. Because the loans are overcollateralized, the protocol itself is protected against the risks of defaults or unexpected falls in the value of the staked currency.

Importantly, another key advantage is that because loans are overcollateralized, if the split of staking fees mostly go towards the protocol rather than the borrower, this leads to higher staking yield than would have been achievable via direct staking (167\% in the example above).  

\section{Stabilization Mechanisms}

But the scenario of always being able to offer savings interest rates as high as staking fees yield is an unrealistic ideal. It assumes that all of the savings deposited by savers will be successfully loaned to borrowers. In reality, borrower demands fluctuate. If the terms of the loans are sufficiently favorable, enough borrowing should happen. It is understandable that borrowers will find a lower overcollateralization ratio attractive, as it allows them to borrow a higher amount with the same collateral size. But not maintaining a high overcollateralization ratio would mean that the protocol is exposed to a higher level of risks, so the room for maneuver in this regard is not unlimited. 

This leaves the other parameter, the ratio for the split of the staking fees, as the primary leverage for regulatory control. In the Anchor protocol, when there is insufficient borrowing, the split changes so that the borrower can keep a relatively higher portion of the staking fee; when enough of the savings are successfully loaned out, this split changes in the other direction so that an increased amount of the staking fee goes towards the savers.

It is possible that depending on the availability of other borrowing opportunities in the market, even at a very favorable split of the staking fees there may not be enough borrowing demand. Some other lenders may offer a lower interest rate; some are already offering 0\% \cite{liquidity}. Others may also be able to offer a lower overcollateralization ratio. We will discuss more about these possible competitions in the next section.

As such, borrower demand will necessarily fluctuate over different market conditions. This is why the current Anchor interest rate for savings is set at a few percentage points below expected staking fees income. This allows the extra income to go into a reserve. So if future staking fees income is to fall, the interest rate for savings does not have to change immediately without sufficient warning to the savers. By using the money from the reserve, the interest rate can be maintained temporarily even if staking fees income cannot sustain the rate in the longer run.

\section{Some Caveats}

At the time of writing, the initially offered 20\% annual rate Anchor interest rate has been maintained around the same level successfully since inception earlier this year (2021). Notably, even during the market crash in May 2021 when the price of Bitcoin dropped by over 50\%, the Anchor rate sustained at around 18\% which was within the expected limit. It has been suggested that this highly competitive rate may become a new industry standard \cite{mansor_2021}. 

However, staking derivatives are a relatively new type of financial instrument. The market is still at an early stage of development, with new trends emerging rapidly. At the time of writing, Anchor only accepts staking derivatives from the native Luna network as collaterals. Although there are plans to accept staking derivatives from other major PoS networks in the near future, it may be more difficult to fully anticipate what that would entail. One reason is that staking derivative tokens generated by staking the native Luna currency are perhaps not so widely accepted as collateral for loans yet. Therefore, holders of these derivative tokens do not have many other options for generating immediate liquidity via borrowing. 

But if this general protocol design is to scale up, and to accept other staking derivatives as collaterals, one has to face the market competition offered by other lenders. As mentioned in the last section, some other lenders may be able to afford taking a higher level of risk, by requiring a relatively marginal overcollateralization ratio as low as just 110\% \cite{liquidity}. At such a tight margin, should the value of the staked currency suddenly drop substantially during a flash crash, one would have to seize the collateral and sell it very quickly to avoid a loss. The Liquity protocol \cite{liquidity}, for instance, is able to do so because of a more efficient and automated liquidation process supported by an internal liquidity pool. Furthermore, they are also able to offer zero interest rate for the borrowers. Currently Liquity only accepts Ethereum (not its staked derivatives) as collateral. But it is conceivable that similar forms of competition may soon come into play for other major cryptocurrencies, as well as their staking derivatives.

According to one analysis \cite{sin_2021}, the Anchor protocol may do particularly well during a bear market. On the savings side, this seems intuitive; saving demands should increase as riskier investment opportunities do not look promising. However, the borrower demands are again more difficult to predict. In a mild downturn, it is possible that some ‘HODLers’ may be more inclined to stake their cryptocurrencies, to wait for the expected market recovery. But if the market is perceived to be in a strong downward trend, or if it is highly volatile, fewer people may be willing to stake. 

Importantly, the main motivation for borrowing may be to create leverage for further investment. In a downturn, the risk of liquidation (as one fails to maintain the overcollateralization ratio, such as during a flash crash) may render this rather unappealing. Accordingly, since May 2021, the Anchor Protocol have had to offer extremely generous staking fee splits to attract borrowing. At such splits, even if borrowing demand is high, the earned fees cannot fully support the high interests paid out to savers. That is because the income of the protocol is just a multiplicative product of total staking derivative deposited as collateral and the fees split earned per collateral unit. As such, in July 2021, external funds had to be injected into the reserve in order to sustain the high savings interest rate \cite{agora_2021}. 

Despite these caveats, like many others \cite{mansor_2021,sin_2021}, we agree that Anchor is a very timely product, with elegant protocol design and stimulating foresight. Below we outline some areas of further development, for achieving similar goals.

\section{Diversification, Interest Rates\\Swaps, and Tranching}

As reviewed so far, the key idea behind the generation of highly competitive interest rates in decentralized finance is to make use of the staking fees as a source of income for savings. In section 4 we mentioned some advantages of not directly staking the savings, and only receiving staking yield via the lending market. However, there may also be advantages in directly engaging with generating income with the deposited savings, rather than lending the money out to borrowers. One such advantage is the economy of scale.
 
For example, the Yearn protocol \cite{yearn} systematically searches for maximal gain with a combination of lending interest rates, staking fees, leveraged reinvestments of the borrowed money, etc , and can achieve much higher return than simple staking strategies. It takes considerable effort to keep track of the changing fee rates, availability of new yield-earning products, etc. But for a decentralized organization at the scale of a bank, the overhead cost incurred is more likely to be worthwhile.

One lucrative way of generating passive income is liquidity pool yield farming \cite{yield_farming}. In decentralized finance, in order to create a market to exchange, rather than relying on a traditional central market maker, one needs to create a pool of funds to allow trades to happen instantly and smoothly. By contributing to such a liquidity pool, one earns fees in return, when transactions take place. In other words, for an automated market to be able to allow trades between two currencies X and Y, that mechanism needs to have enough flowing infantry of both currencies. But merely putting down these currencies into the mechanism, one is thereby facilitating trades. So, in a way not unlike staking, one earns the deserved reward through the process; accordingly this kind of ‘investment ‘ is sometimes called liquidity pool ‘staking’, even though it is distinct from the type staking in PoS networks mentioned earlier. 

The fees one can earn in joining liquidity pools are typically on the order of under 10\% annual rate. However, when a new pool is formed, often there are extra incentives to attract initial endowment, and those can be extremely lucrative, sometimes reaching over 100\% annual rate. The process of earning these high annual rates is sometimes called yield farming, as one is essentially investing into a new pool hoping that it would grow as expected, to allow yields to be ‘harvested’.

As expected, at such high annual rates there are also risks involved. The incentives in yield farming are typically given in the native currency of the protocol, which may not turn out to be so valuable in the long run if the protocol does not turn out to be successful. Besides that, similar to staking, the value of the assets deposited into the liquidity pool may fluctuate in time. Besides simple depreciation, when the balance of the two deposited assets changes (e.g. one increases in value while the other does not), this can create another kind of deficit known as impermanent loss \cite{xu2021sok,aigner2021uniswap}.

Despite these risks, there are arguments to be made that as a bank-like organization, these are investment opportunities worth pursuing. One reason is that at a large enough scale, one can methodically and effectively make use of insurance and hedging instruments. For example, earlier in section 3 we mentioned the use of interest rate swaps to turn a variable-rate future income into fixed-rate income. In decentralized finance, protocols like Horizon \cite{horizon} are also creating opportunities for doing so. Even in the absence of a fluid interest rates market, Horizon makes use of game theoretic and auction-like mechanisms to facilitate the swaps. This way, one can opt to take a smaller but fixed yield from farming liquidity pools.

To manage risks, protocols like Barnbridge \cite{barnbridge} and Saffron \cite{saffron} also allow one to break down an investment or loan into ‘tranches’, each representing different risk levels. For example, a ‘senior’ tranche may lead to a lower yield overall, but the rate will be more guaranteed. A bank-like organization can select the appropriate risk level given the reserve level at the moment, with the guiding principle that higher risks are only affordable when there is sufficient excess in reserve. 

These strategies are not mutually exclusive with extracting staking fees via borrowers’ collateral, which is admittedly virtually risk-free. That is because when there is enough borrowing demand, essentially the borrowers absorb most of the risk involved. However, these riskier and higher yield strategies can be employed in parallel, especially when borrowing demand is low. As with competitive banks in the real world, financial institutions typically participate in various activities including both investment and lending. If the goal is to give savers the highest stable interest rates at minimal risks, an algorithm that explicitly optimizes for the ideal combination of different strategies to achieve this goal should be in principle more advantageous than fixating onto a single strategy a priori.

\section{Towards Universal Basic Income}

We started by considering the high stable interests of decentralized finance protocols from the perspective of common household savers. Currently, few of them have access to investment opportunities with such stable high yield. However, the availability of these products will naturally also be accessible by the wealthier, seasoned investors too. Therefore, are they actually going to help the common household savers, or are they just going to make the rich richer still?

There are perhaps reasons to be optimistic (from an egalitarian perspective). The first is that even if these products do allow the wealthier to benefit, the margin of this benefit is unlikely to be as high as that for the common savers. That is because currently the savvy and capital-rich investors already have access to various high return investment options, including high-entry, actively managed hedge funds. 

Secondly, with social and keyless wallets \cite{one_wallet}, one ultimate goal is to allow everyday smartphone users to be able to access decentralized finance applications securely with ease. For many of these new adopters, having access to such high stable return investment options will likely help level the playing field.

However, we must also acknowledge that it is not easy to foresee a future where such high-yield, low-risk products will become commonplace. The impact of this for traditional financial products will lead to a chain of consequences that would ultimately depend on relatively unpredictable factors, including possible governmental intervention. Accordingly, we cannot expect that the lofty vision of universal basic income will naturally materialize out of market forces. Rather, we should see the availability of these new decentralized finance products as providing opportunities for such an agenda. 

For example, there are already projects in the cryptocurrency space which aim at providing universal basic income to those who can prove their identity \cite{proof_humanity}. Capitalizing on the high yield generated by the Anchor, the Angel protocol \cite{angel_protocol} also allows charity donations to become virtually perceptual (as supported by the future yields). These may provide the platform for some universal basic income schemes to be implemented. 

As a longer-term vision, radical markets \cite{posner2019radical} on blockchain can also help to enforce overall fairness. Because of the permanent and immutable nature of transaction records in cryptocurrency ecosystems, taxation will eventually become automatically enforced, with evasion becoming extremely difficult. Importantly this can also open up venues for novel and interesting ways of redistributing wealth. For example, under a Harberger taxation scheme \cite{posner2019radical}, the value of a property can be self-assessed, i.e. determined by the owner. The owner may be incentivized to not to overvalue the property, in order to minimize taxation. At the same time, transaction of the ownership can be made mandatory if any buyer offers a price above the self-assessed value. Therefore the owner should also not undervalue the property. With a well-defined true value to the owner, properties can be fluidly and fairly transferred. Economic analyses have shown that this can motivate and orchestrate the entire community into maximizing common good, while maintaining market efficiency. Similar mechanisms can also be implemented to implement highly progressive and quadratic wealth redistribution. 

Importantly, even in the shorter-term, there is also an argument to be made that allowing the rich to become richer should not always be a concern, even for equalitarian purposes. An individual’s fixed or basic income may be dwarfed by the wealth of others, if such income is paid in a currency that is subject to inflation. If the rich can offer higher prices for goods, amid competition there may be a concern that one’s income may lose purchasing power. However, this would not be an issue if the basic income is to be paid out in a currency that is truly resistant to inflation. The US dollar is supposedly (loosely) pegged to the prices of goods. However, there has been criticism that the consumer price index (CPI) provided by the government is biased towards goods that do not truly reflect the quality of life of an average individual \cite{ssa}. Currently, there are projects such as Frax’s stable peg and index which aim to rectify this situation \cite{olympus,frax,reflexer}, to create a stable currency in decentralized finance pegged not to the inflation-prone US dollar, but to a more meaningful representation of the true cost of living. 

Imagine if a stable yield of as high as over 10\% per annum is to be paid out in this inflation-resistant currency. With such a product, to support the universal basic income for a person’s lifetime, one would need a few times the individual’s annual living cost, as a one-time endowment. This is regardless of the future conditions of the economy, or how the wealthy may become richer - if the currency truly reflects the cost of living at the moment, in a stable peg. This could be the basis on which a meaningful universal basic income can be made feasible.

\section{Closing Remarks}

We have reviewed the logic of how the generation of a stable savings interest rate at as high as 20\% annual rate is possible. In traditional financial investments, such a high and stable rate is above what even some of the best investors can achieve these days \cite{ro_2021}. The key innovation here is to capitalize on the unique features of decentralized networks and lending marketplaces.  In particular, ordinary savers can participate in multiple network stakings and fixed-rate saving products.

It is important to recognize that this is not a technical sleight of hand. The economics behind it are straightforward. Through staking, one contributes to a process that allows highly efficient and robust financial services and products to be delivered, along with other applications \cite{miller_2021,davinci} of the blockchain system. Outside of the context of network staking, the same concept applies for liquidity pools, which allow automated decentralized markets to take place. To capture the growth of the cryptocurrency market, one can also turn these into relatively fixed-rate future incomes via interest rate swapping mechanisms, and thereby use the income to further finance stable income for the savers.

In the traditional financial world, some bankers may lose part of their bonuses during economic downturns. But it rarely cuts into their base salary level, because their services are supposedly still valued. Likewise, even as the cryptomarket sees adjustments, staking fees in decentralized networks are here to stay, for staking contributions to the automation of jobs akin to those done by bankers and corporate lawyers. The very premise of decentralized finance is that everybody can participate fairly in the market mechanisms. Unlike in the traditional financial system, in decentralized finance everyone can be a ‘banker’ – even if only passively, indirectly, on a smaller scale. This is the economic basis of how we can possibly provide the household savers with meaningful basic wealth.

\section{Acknowledgements}

We thank Sahil Dewan, Giv Parvaneh, and Boris Polania for helpful discussions and comments. HL has previously received consulting fees and research funding from Harmony.

\bibliographystyle{unsrt}
\bibliography{references.bib}

\end{document}